\documentclass{lmcs}

\usepackage{enumerate}
\usepackage{hyperref}
\usepackage{amssymb,xcolor,verbatim,soul}

\theoremstyle{plain}
\newtheorem{example}[thm]{Example}

\newcommand\twoheaduparrow{\mathrel{\rotatebox{90}{$\twoheadrightarrow$}}}
\newcommand\twoheaddownarrow{\mathrel{\rotatebox{90}{$\twoheadleftarrow$}}}
\def\kurung#1{\left(#1\right)}
\def\himp#1{\{#1\}}

\def\real#1{\mathbb{#1}}

\def\mc#1{\mathcal{#1}}
\def\sub{\subseteq~\!\!\!}

\def\thdai{\twoheaddownarrow_{\Irr}}
\def\lli{\ll_{\Irr}}
\def\dua{\twoheaduparrow_{\Irr}}
\def\siconv{\xrightarrow{\Irr}}

\newcommand{\intr}{\operatorname{int}}
\newcommand{\cl}{\operatorname{cl}}

\newcommand{\Irr}{\operatorname{Irr}}
\newcommand{\SI}{\operatorname{SI}}

\newcommand{\bigsup}{\bigvee}

\newcommand*{\affmark}[1][*]{\textsuperscript{#1}}
\begin{document}

\title[On a new convergence class in sup-sober spaces]{On a new convergence class in sup-sober spaces}


\author[H. Andradi]{Hadrian Andradi \affmark[1,2]}	
\author[W. K. Ho]{Weng Kin Ho \affmark[1]}	
\address{\affmark[1] National Institute of Education, Nanyang Technological University, 1 Nanyang Walk, Singapore 637616}	
\address{\affmark[2] Department of Mathematics, Faculty of Mathematics and Natural Sciences, Universitas Gadjah Mada, Indonesia 55281}	
\email{hadrian.andradi@gmail.com (H. Andradi),  wengkin.ho@nie.edu.sg (W.K. Ho)}  
\thanks{The first author is supported by Nanyang Technological University Research Scholarship (RSS)}	




\begin{abstract}
 \noindent Recently, J. D. Lawson encouraged the domain theory community to consider the scientific program of developing domain theory in the wider context of $T_0$-spaces instead of restricting to posets.  In this paper, we respond to this calling by proving a topological parallel of a 2005 result due to B. Zhao and D. Zhao, i.e., an order-theoretic characterisation of those posets for which the Scott-convergence is topological.  We do this by adopting a recent approach due to D. Zhao and W. K. Ho by replacing directed subsets with irreducible sets. As a result, we formulate a new convergence class $\mc{I}$ in $T_0$-spaces called $\Irr$-convergence and establish that a sup-sober space $X$ is $\SI^{-}$-continuous if and only if it satisfies $*$-property and the convergence class $\mc{I}$ in it is topological.
\end{abstract}

\maketitle

\section{Introduction}
\label{sec: intro}
Domain theory can be said to be a theory of approximation on partially ordered sets.
There are two sides of the same domain-theoretic coin: the order-theoretic one and the topological one.  On the order-theoretic side, the facility to approximate is built in the ordered structures via approximation relations, and here domain is the generic term that includes all ordered structures that satisfy some approximation axioms.  On the topological side, approximation can be handled by \emph{topology}; more precisely, using net convergence.  Two famous results of D. S. Scott \cite{scott72a} epitomise this deep connection between domains and topology:  (1) A space is injective if and only if it is a continuous lattice with respect to its specialization order. (2) The Scott-convergence class in a directed complete partial order (dcpo, for short) $P$ is topological if and only if $P$ is continuous (furthermore, in a continuous dcpo, the Scott topology induces the Scott convergence). The second result was later generalised by B. Zhao and D. Zhao (\cite{zhaozhao05}) to the setting of posets which are not necessarily dcpo's. We highlight to the reader that, in \cite{zhaozhao05}, the terminology ``lim-inf convergence'' is used instead of ``Scott-convergence''. The latter seems more suitable to use since the former is a bit misleading, because in \cite{gierzetal03}, which is the modern dominant source for Domain Theory, lim-inf convergence is more related to the Lawson than the Scott topology (see \cite[Theorem III-3.7]{gierzetal03}). The terminology ``Scott-convergence'' was used in \cite{erne81} in which the fact that ``a poset $P$ is continuous if and only if all the sets $\twoheaddownarrow x$ are directed and the Scott-convergence class in $P$ is topological'' is proved  by considering filter convergence rather than net convergence (see \cite[Theorem 2.13]{erne81}). 

In an invited presentation\footnote{This talk bears an extra-terrestrial title of ``\emph{Close Encounters of the Third Kind: Domain Theory Meets $T_0$-Spaces Meets Topology}".} at the 6th International Symposium in Domain Theory, J. D. Lawson gave further evidence from recent development in domain theory to illustrate this intimate relationship between domains and $T_0$-spaces.  In particular, it was pointed out that ``several results in domain theory can be lifted from the context of posets to $T_0$-spaces".  For example, (1) the topological technique of dcpo-completion of posets~\cite{zhaofan07} can be upgraded to yield the D-completion of $T_0$-spaces (i.e., a certain completion of $T_0$-spaces to yield d-spaces) \cite{keimellawson09}, and (2) an important order-theoretic result known as Rudin's lemma~\cite{gierzetal83}, which is central to the theory of quasicontinuos domains, has a topological version \cite{heckmannkeimel13}.

In this paper, we respond (in a small way) to Lawson's call to develop the core of domain theory directly in topological spaces by establishing a topological parallel of the aforementioned result due to B. Zhao and D. Zhao (\cite[Theorem 2.1]{zhaozhao05}). To prove a parallel topological result of this, we adopt the recent approach in~\cite{zhaoho15} by replacing directed subsets with irreducible subsets.  The motivation for their approach is based on the observation that the directed subsets of a poset are precisely its Alexandroff irreducible subsets.
Based on this replacement principle, we invent topological analogues of the usual domain-theoretic notions:
(i)  a new way-below relation $\ll_{\Irr}$ on a $T_0$-space,
(ii) some new notions of continuity of spaces, and
(iii) a new net convergence class $\mathcal{I}$ on a given topological space $X$.

Working with the so-called irreducible-directed replacement principle have a connection with the concept of subset system $\mc{Z}$ introduced in \cite{wrightwagnerthatcer78}. The $\mc{Z}$-theory in partially ordered sets have been studied extensively in the last few decades (see, e.g., \cite{bandelterne83}, \cite{baranga96}, \cite{erne99}, \cite{venogupalan86}, \cite{xuliu03}). In light of this theory, the replacement principle here may be saved for a particular subcollection $\mc{Z}$ in the realm of $T_0$-space.  Moreover, this generalization would correlate the theory of $\mc{Z}$-(quasi)continuous posets and their topological aspects.

In this paper, the notion of sup-sobriety is heavily involved. Thi notion, which was first introduced in \cite{zhaoho15} as a generalisation of bounded-sobriety (\cite{mislove99}), has close connections with irreducibly-derived topology mentioned in \cite{zhaoho15}.  Because little is known about this kind of sobriety, it is one of the purposes of this paper to give a slightly better understanding of it in relation to net convergence.

We organise this paper in the following way. In Section~\ref{sec: prelim}, we summarise some of the recent results reported in~\cite{zhaoho15} that are essential in our ensuing development. These results concern the irreducibly-derived topology defined using irreducible sets of the underlying topology $X$ and sup-sober spaces. In Section~\ref{sec: irr-cont spaces}, we focus in some result in some continuities of a space. In Section~\ref{sec: convergence class defined by irreducible sets}, we introduce the new convergence class $\mc{I}$ defined in any $T_0$-space $X$ and present some of its elementary properties.  Finally, we focus our development of the convergence class $\mc{I}$ on sup-sober spaces and prove the main characterisation theorem which we advertised in the abstract.

\section{Irreducibly derived topology}
\label{sec: prelim}
A nonempty subset $E$ of a topological space $(X,\tau)$ is \emph{irreducible} if for any closed sets $A_1$ and $A_2$,
whenever $E \subseteq A_1 \cup A_2$, either $E \subseteq A_1$
or $E \subseteq A_2$.  The family of all irreducible subsets of $X$ is denoted by $\Irr_{\tau}(X)$ or $\Irr(X)$ whenever it is clear which topology one is referring to.

It is often useful to check the irreducibility of a set using open sets, i.e., a nonempty set $E$ is irreducible if and only if for any open sets $U_1$ and $U_2$, if $E \cap U_i \neq \emptyset$ ($i = 1,2$), then $E \cap U_1 \cap U_2 \neq \emptyset$.  Regarding irreducible sets, here are some elementary properties:
\begin{prop}
For any given topological space $(X,\tau)$, one has:
\begin{enumerate}[\em(1)]
\item $E \in \Irr_{\tau}(X)$ if and only if $\cl(E)\in \Irr_{\tau}(X)$.
\item The continuous image of an irreducible set is again irreducible.
\item If $\nu$ is some other topology on $X$ with $\nu\sub\tau$, then $\Irr_{\tau}(X)\sub\Irr_{\nu}(X)$.
\end{enumerate}
\end{prop}

Every $T_0$-space $(X,\tau)$ can be viewed as a partially ordered set via its \emph{specialisation order}, denoted by $\leq_\tau$, where $x \leq_{\tau} y$ if $x \in \cl_{\tau}(y)$. Henceforth, all order-theoretical statements on a $T_0$-space refer to the specialisation order on the space.  For any subset $A$ of a $T_0$-space $(X,\tau)$, the supremum of $A$, denoted by $\bigsup_{\tau} A$, is the least upper bound of $A$ with respect to the specialisation order $\leq_\tau$ of $X$. We denote the set of all irreducible subsets of $X$ whose supremum exists by $\Irr_{\tau}^{+}(X)$. The subscript ``$_\tau$'' shall be removed from the denotations whenever it is clear which topology one is referring to.

A topological space $X$ is \emph{sober} if every irreducible closed set is the closure of a unique singleton.  All Hausdorff spaces are sober and all sober spaces are $T_0$.  The Scott space of any continuous domain is sober.  A weaker form of sobriety is that of \emph{bounded-sobriety} which requires that every irreducible closed set which is bounded above with respect to the specialisation order is the closure of a unique singleton. Notice that, as a specialisation order is involved, a bounded sober space needs to be $T_0$ at the first place. Bounded-sober spaces have been studied in~\cite{mislove99} and~\cite{zhaofan07}. A yet weaker form of sobriety is that of sup-sobriety.  A $T_0$-space is \emph{sup-sober} if every closed set $F \in \Irr^+(X)$ is the closure of a unique singleton, in this case $F$ is exactly $\cl\{\bigvee F\}$. Every $T_1$-space is sup-sober. Every poset $P$ is sup-sober with respect to its upper topology, i.e., the coarsest one generated by sets of the form $P \backslash \downarrow x$,~$x \in P$. All continuous posets are sup-sober with respect to the Scott topology, yet a sup-sober space is not necessarily continuous as witnessed by Johnstone's space~(\cite{johnstone81}).

Directed subsets play a central role in domain theory.  Directed subsets of a poset can be characterised topologically.  Recall that the Alexandroff topology on a poset $P$ consists of all upper sets.  The directed subsets of $P$ are precisely the Alexandroff irreducible subsets. The Scott topology is a coarsening of the Alexandroff topology in that every Scott open set is required to be an upper set and in addition inaccessible by directed suprema.  By replacing the directed sets by irreducible sets in the definition of a Scott open set, D. Zhao and W. K. Ho defined for any $T_0$-space (not just poset) a coarser topology called the \emph{irreducibly-derived topology} that mimics the Scott topology on a poset.  More precisely, let $(X,\tau)$ be a $T_0$-space and $U \subseteq X$, define $U \in \tau_{SI}$ if
\begin{enumerate}[(1)]
\item $U \in \tau$, and
\item for every $E \in \Irr_{\tau}^{+}(X)$, $\bigsup E \in U$ implies $E \cap U \neq \emptyset$.
\end{enumerate}
It can be easily verified that $SI(X,\tau) := \kurung{X,\tau_{SI}}$ is a topological space whose topology is coarser than $(X,\tau$). An open set in $SI(X,\tau)$ is called $SI$-open and the interior of a subset $A$ of $X$ with respect to $\tau_{SI}$ is denoted by $\intr_{SI}(A)$.

Because the Scott-like topology $\tau_{SI}$ is derived from a topology $\tau$ on the same set $X$, we sometimes refer to $\tau_{SI}$ as the \emph{Scott derivative} of $\tau$.

\begin{prop} \cite{zhaoho15}
Let $(X,\tau)$ be a $T_0$-space. Then the following hold:
\begin{enumerate}[\em(1)]
\item The specialisation orders of spaces $(X,\tau)$ and $SI(X,\tau)$ coincide. 
\item A closed subset $C$ of $(X,\tau)$ is closed in $SI(X,\tau)$ if and only if for every $E \in\Irr_{\tau}^{+}(X)$, $E \subseteq C$ implies $\bigsup E \in C$.
\item An open subset $U$ of $(X,\tau)$ is $SI$-open if and only if for any closed set $E \in \Irr_{\tau}^{+}(X) $, $\bigsup E \in U$ implies $E \cap U \neq\emptyset$.
\item A subset $U$ of $X$ is clopen in $(X,\tau)$ if and only if it is clopen in $SI(X,\tau)$.
\item $(X,\tau)$ is connected if and only if $SI(X,\tau)$ is connected.
\end{enumerate}
\end{prop}

\begin{example}\label{ex:si_topology} Let $P$ be a poset endowed with the Alexandroff topology $\alpha(P)$. Since the irreducible sets in $(P,\alpha(P))$ are precisely the directed ones, it is clear that $SI(P,\alpha(P)) = \Sigma(P)$, where $\Sigma(P)$ is the set $P$ endowed with the Scott topology on $P$.
\end{example}

In general, the Scott topology of a given poset does not coincide with its Alexandroff topology.   For example, in the set $\real{R}$ of all real numbers equipped with the usual order,  sets of the form $[x,\infty)$ are Alexandroff open but not Scott open. We shall now look at those spaces which are equal to their Scott derivatives. A $T_0$-space $(X,\tau)$ is said to satisfy $SI^{\infty}$\emph{-property} if $\tau = \tau_{SI}$. 

Given $T_0$-space $(X,\tau)$, one can derive from it a space satisfying $SI^{\infty}$-property.
Let $(X,\tau)$ be a $T_0$ space and $\alpha$ an ordinal.
We define by transfinite induction a topological space $X^{\alpha}$ on $X$ as follows:

\begin{enumerate}[(1)]
\item $X^{0} := (X,\tau)$;
\item $X^{\alpha + 1} := SI\kurung{X^{\alpha}}$;
\item If $\alpha$ is a limit ordinal, then $X^{\alpha}$ is the space on $X$ whose topology is the intersection of all topologies $X^\beta$, where $\beta < \alpha$.
\end{enumerate}
Since $(X_\alpha)_\alpha$ is a sequence of increasingly coarser
topologies on $X$, there is a smallest ordinal $\gamma(X)$ such that the topology on $X^{\alpha}$ coincides with that on $X^{\gamma}$ for all $\alpha\geq \gamma(X)$. We denote this $X^{\gamma(X)}$ by $X^{\infty}$.

It is immediately clear by the definition that the following theorem holds.
\begin{thm} \cite[Theorem 4.5]{zhaoho15}\label{th:SI-infty}
\label{thm: char of kb-space}
A $T_0$-space $X$ is sup-sober if and only if it satisfies $SI^{\infty}$-property.
\end{thm}

\section{Some continuities of spaces}
\label{sec: irr-cont spaces}
Starting from this section, a topological space or a space refers to a $T_0$-space, unless otherwise mentioned.

In a space $X$, one defines a ``new" way-below relation $\lli$ (called the \emph{$\Irr$-way-below relation}) using irreducible subsets instead of directed subsets.
Given $x, y \in X$, the $\Irr$-way-below relation is defined as follows:
\[
x \lli y \iff \forall E \in \Irr^{+}(X).~(\bigsup E \geq y) \implies (E ~\cap \uparrow x \neq \emptyset).
\]
For a given $x \in X$, $\thdai x$ denotes the set $\himp{y \in X \mid y \lli x}$.
The following properties of $\Irr$-way-below relation are as expected:
\begin{prop}
\label{prop: transitivity of lli}
In a space $X$ the following hold for all $u,x,y$ and $z \in X$:
\begin{enumerate}[\em(1)]
  \item $x \lli y$ implies $x \leq y$.
  \item $u \leq x \lli y\leq z$ implies $u \lli z$.
  \item $x\in \intr_{SI}(\uparrow y)$ implies $y\lli x$.
\end{enumerate}
\end{prop}

Using $\lli$, we can now introduce the notion of $\Irr$-continuous space -- a topological analogue of continuous posets.
\begin{defi}
A space $X$ is said to be $\Irr$\emph{-continuous} if for every $x\in X$ the following hold:
\begin{enumerate}[(1)]
\item $\thdai x$ is irreducible and
\item $x = \bigsup \thdai x$.
\end{enumerate}
\end{defi}
\begin{rem}
Our definition of $\Irr$-continuous space differs from that of $\SI$-continuous spaces defined in \cite[p.192]{zhaoho15} in that we choose to drop their first condition, i.e., for any $x \in X$, the set $\dua x:=\himp{y\in X\mid x\lli y}$ is open in $X$, and weaken the requirement in their second condition: from $\thdai x$ being directed to $\thdai x$ being irreducible.
One also needs to notice that sticking in the definition of $\SI$-continuity from \cite[p.192]{zhaoho15} will go contrary to our original intention of developing domain theory in the wider contexts of topological spaces and \emph{not restricted just to (continuous) posets}. This is because of a result by M. Ern\'{e} (\cite[Theorem 4, p.462]{erne05}).  That result asserts that a topological space is a weak C-space (i.e., it is both a C-space and a weak monotone convergence space) if and only if it is homeomorphic to the Scott space of some continuous poset.  It was shown in~\cite[Theorem 6.4]{zhaoho15} that $X$ is $\SI$-continuous if and only if the derived topology $SI(X)$ is a C-space.  Because $SI(X)$ is always a weak monotone convergence space, it follows that the derived topology on an $\SI$-continuous space is homeomorphic to the Scott topology on some continuous poset.
\end{rem}

With the absence of the first condition and weakened version of second condition, we can still say a few things about $\Irr$-continuous spaces in general.

\begin{lem}
\label{lem: prelude to interpolating}
Let $X$ be an $\Irr$-continuous space.  Then, for every $x \in X$ it holds that
\[
x=\bigsup \bigcup \himp{\thdai y\mid y\lli x}.
\]
\end{lem}
\proof
Let $M_x := \bigcup \himp{\thdai y\mid y\lli x}$. It is clear that $x$ is an upper bound of $M_x$. Let $u$ be an upper bound of $M_x$. We shall show that $u \geq x$ for any upper bound $u$ of $M_x$. Suppose for the sake of contradiction that $u \ngeq x$.  Then, by the $\Irr$-continuity of $X$, $x = \bigsup \thdai x$ so that there exists $y \in ~\thdai x$ with $y \nleq u$.  Repeating the same argument we can find a $z \in ~\thdai y$ such that $z \nleq u$.  But this is a contradiction to the fact that $z \in M_x$ and $u$ is an upper bound of $M_x$.  Therefore, $u \geq x$ and this completes the proof.\qed

Any domain theorist would know the price for weakening the second condition, i.e., one loses the \emph{interpolating property} of the Irr-way-below relation.  Fortunately, within the scope of our present study concerning sup-sober spaces, we can recover this loss.

\begin{thm}\label{thm: interpoting prop in SI infty}
Let $X$ be an $\Irr$-continuous and sup-sober space.
Then, $\lli$ enjoys the interpolating property in that whenever $z\lli x$, there exists $y \in X$ such that
\[
z \lli y\lli x.
\]
\end{thm}
\proof
We first show that $M_x := \bigcup \himp{\thdai y\mid y\lli x}$ is an irreducible subset of $X$. Let $U_1$ and $U_2$ be open  in $X$ such that $M_x\cap U_1\neq\emptyset$ and $M_x\cap U_2\neq\emptyset$. Then there exist $y_1,y_2\in ~\thdai x$ such that $y_1\in U_1$ and $y_2\in U_2$. Since $x$ is an upper bound of $\himp{y_1,y_2}$ and both $U_1$ and $U_2$ are upper sets, $x\in U_1\cap U_2$.  By $\Irr$-continuity of $X$, $x$ is the supremum of $\thdai x$. Since $X$ is sup-sober, it enjoys the $SI^{\infty}$ property and so $U_1,~U_2\in SI(X)$.  Hence there exists $y \in~ \thdai x$ such that $y \in U_1 \cap U_2$. Using a similar argument, there exists $z \in~ \thdai y$ such that $z \in U_1 \cap U_2$. Therefore, there exists $z \in X$ such that $z \in M_x \cap U_1 \cap U_2$. Consequently, $M_x$ is an irreducible subset of $X$.  Now, let $z \lli x$.  Since $M_x$ is irreducible and, by Lemma~\ref{lem: prelude to interpolating}, $\bigsup M_x = x$, there exists $w \in M_x$ such that $z \leq w$. Hence there exists $y \in X$ such that, by virtue of Proposition~\ref{prop: transitivity of lli}, $z \lli y \lli x$ holds as desired. \qed

\begin{example}
The rational line $Q := (\mathbb{Q},\leq)$ with the Scott topology $\Sigma Q$ is an $\Irr$-continuous sup-sober space which is not sober.
\end{example}

Another way to recover the interpolating property of the $\Irr$-way-below relation is by considering $\SI$-continuity introduced in \cite{zhaoho15} but omitting the first condition. We define $\SI^-$-continuity of spaces as follows:
\begin{defi}
A space $X$ is said to be \emph{$\SI^-$-continuous} if for every $x\in X$ the set $\thdai x$ contains a directed set whose supremum is $x$.
\end{defi}

From the definition, one can see that every $\SI^-$-continuous space is $\Irr$-continuous. This is because in any space, directed sets are irreducible. In particular, in any poset endowed with the Alexandroff topology on it, both notions of continuity are exactly the same.

An $\SI^-$-continuous space and $\Irr$-continuous space shares a same property: For every point $x$, $\thdai x$ contains an irreducible subset whose supremum is $x$. Unlike on an $\Irr$-continuous space, the $\Irr$-way-below relation on an $\SI^-$-continuous space is interpolating, regardless of whether it is sup-sober.

\begin{prop}\label{prop:interpolation_in_SI*}
On an $\SI^-$-continuous space, the relation $\lli$ enjoys the interpolating property.
\end{prop}
\proof
Let $X$ be an $\SI^-$-continuous space and $z,x\in X$ such that $z\lli x$. By definition, there exists $D\sub ~\thdai x$ such that $D$ is directed and $\bigvee D=x$. For each $d\in D$, we fix a directed subset $B_y$ of $\thdai y$ whose supremum is $y$. Now consider the set $A=\bigcup\himp{B_y\mid y\in D}$. Being a union of directed sets, $A$ is directed, hence irreducible. The fact that $X$ is $\SI^-$-continuous gives $\bigvee A=x$. We then have an element $y^\prime\in A$ such that $z\leq y^\prime$. This gives the existence of an element $y\in X$ such that $z\lli y\lli x$. \qed 

An $\SI^-$-continuous sup-sober space satisfies a special property, that is, for every $F\in\Irr^+(X)$ there exists a directed subset $D$ of $\downarrow F$ such that $\bigvee F=\bigvee D$. We call such property \emph{$*$-property}. It is given in \cite[Lemma 7.4]{zhaoho15} that every C-space satisfies $*$-property. Recall that a $T_0$-space is a \emph{C-space} if for every open set $U$ and $x\in U$ there exists $y\in U$ satisfying $x\in\intr(U)$.

\begin{prop}\label{prop:SI-+kb_implies_*property}
\hfill
\begin{enumerate}
\item Every $\Irr$-continuous space satisfying $*$-property is $\SI^-$-continuous.
\item Every $\SI^-$-continuous sup-sober space satisfies $*$-property.
\end{enumerate}
\end{prop}
\proof \hfill
\begin{enumerate}
\item The proof is immediate from the definition.

\item 
Let $X$ be a $\SI^-$-continuous sup-sober space and $F\in\Irr^+(X)$. We then, by sup-sobriety of $X$, have $\cl(F)=\cl(\himp{x})$ where $x=\bigsup F$. This implies $\thdai x\sub~ \downarrow F$. The fact that $X$ is $\SI^-$-continuous implies that there exists a directed subset $D$ of $\downarrow F$ whose supremum is $x$, as desired. \qed
\end{enumerate}

At the end of this section, we shall present some results concerning continuities of spaces. We first recall the definition of $\SI$-continuous.

\begin{defi}
A space $X$ is called \emph{$\SI$-continuous} if it is $\SI^-$-continuous and it satisfies \emph{$\oplus$-property}, i.e., $\twoheaduparrow_{\Irr} x:=\himp{y\in X\mid x\lli y}$ is open in $X$ for each $x\in X$. 
\end{defi}

In the definition of $\Irr$-continuous, one can see that there is no much information about the underlying topology. Imposing $\oplus$-property to a space may give us more information regarding the topology on it. We shall call $\Irr$-continuous space satisfying $\oplus$-property an \emph{$\Irr^+$-continuous space}.

\begin{example}\label{ex:cofinite_topology}
The space $\mathbb{N}$ endowed with a cofinite topology is a $T_1$-space. Hence it is $\Irr$-continuous and sup-sober, yet it does not satisfy $\oplus$-property. This space is also obviously not a C-space.
\end{example} 

It is mentioned in \cite[Theorem 6.4]{zhaoho15} that a space $X$ is $\SI$-continuous if and only if the space $SI(X)$ is a C-space. We shall show that, in the presence of sup-sobriety, the notions of $\SI$-continuity and $\Irr^+$-continuity are the same. In fact, for a sup-sober space, satisfying one of the two continuities is equivalent with being a C-space.

\begin{lem}\label{lem:SI(X)_is_C_then_X_is_Irr+}
If $X$  is a space such that $SI(X)$ is a C-space, then $X$ is $\Irr^+$-continuous.
\end{lem}
\proof
For each $x\in X$, we define $S_x=\himp{y\in X\mid x\in \intr_{SI}(\uparrow y)}$. We first show that $\bigsup S_x=x$. For each $y\in S_x$ we have $x\in \uparrow y$, hence $y\leq x$. Now let $x\nleq z$. Since $SI(X)$ is a C-space and $X-\downarrow z$ is $SI$-open, there exists $y_0\in X-\downarrow z$ such that $x\in\intr_{SI}(\uparrow y_0)$. We have that $y_0\in S_x$ and $y_0\nleq z$. Therefore $\bigsup S_x=x$.

We next show that $S_x$ is irreducible. Let $U_1$ and $U_2$ be opens in $X$ such that $S_x\cap U_1\neq \varnothing$ and $S_x\cap U_2\neq\varnothing$. There exist $y_1\in U_1$ and $y_2\in U_2$ such that $x\in \intr_{SI}(\uparrow y_1)$ and $x\in \intr_{SI}(\uparrow y_2)$, hence $x\in\intr_{SI}(\uparrow y_1) \cap \intr_{SI}(\uparrow y_2)$. Since $SI(X)$ is a C-space, there exists $y_3\in \intr_{SI}(\uparrow y_1) \cap \intr_{SI}(\uparrow y_2)\sub~ \uparrow y_1\cap\uparrow y_2\sub U_1\cap U_2$ such that $x\in\intr_{SI}(\uparrow y_3)$. We have $y_3\in S_x\cap U_1\cap U_2\neq\varnothing$. Hence $S_x$ is irreducible in $X$.

If $y\in S_x$, then $x\in \intr_{SI}(\uparrow y)$. By Proposition \ref{prop: transitivity of lli} we have that $y\lli x$. Thus $S_x\sub~ \thdai x$, implying that $\downarrow S_x\sub~\thdai x$. Now let $y^\prime\lli x$. Since $S_x$ is irreducible and $x\leq \bigsup S_x$, there exists $y\in S_x$ such that $y^\prime\leq y$. Hence $y^\prime\in\downarrow S_x$. Therefore $\downarrow S_x=\thdai x$. At this point, we have that $X$ is $\Irr$-continuous. 

Now if $z\in~\twoheaduparrow_{\Irr}(x)$, then $x\in~\downarrow S_z$. There exists an element $y\in X$ such that $x\leq y$ and $z\in\intr_{SI}(\uparrow y)$. Hence $z\in\bigcup_{x\leq y}\intr_{SI}(\uparrow y)$. If $z\in\intr_{SI}(\uparrow y)$ for some $y\in \uparrow x$, we have that $x\leq y\lli z$. Thus $z\in\twoheaduparrow_{\Irr} x$. We have that
$$\twoheaduparrow_{\Irr} x=\bigcup_{y\in\uparrow x}\intr_{SI}(\uparrow y)$$
which gives $\twoheaduparrow_{\Irr} x$ is open, in particular irreducibly open. \qed

One needs to to notice that the condition $SI(X)$ is a C-space in Lemma \ref{lem:SI(X)_is_C_then_X_is_Irr+} cannot be replaced by $X$ is a C-space. Indeed, there is a C-space which is not $\Irr$-continuous, let alone $\Irr^+$-continuous.

\begin{example}\label{ex:C-space_but_not_Irrcts}
Let $T=\himp{\top,\bot,a,1,2,\ldots}$ equipped with the partial order defined as follows: $\bot\leq x\leq \top$ for all $x\in T$ and $1\leq2\leq \ldots$. Consider the space $T$ endowed with Alexandroff topology. It is clearly a C-space. Notice also that the space is not $\Irr$-continuous since $\thdai a=\himp{\bot}$.
\end{example}

\begin{lem}\label{lem:kbsober+irr+implies_Cspace}
If $X$ is an $\Irr^+$-continuous sup-sober space, then $X$ is a C-space. 
\end{lem}
\proof
We first show that given $x\lli y$, we have $y\in\intr_{SI}(\uparrow x)$.  Let $x\lli y$. Since the relation $\lli$ is interpolating (in light of Theorem \ref{thm: interpoting prop in SI infty}), there exist $z_1,z_2,\ldots, z_n,\ldots\in X$ such that
$$x\lli\ldots \lli z_n \lli \ldots \lli z_2\lli z_1\lli y$$
By assumption, we have that the set $V:=\bigcup_{i\in\mathbb{N}}\twoheaduparrow_{\Irr}z_i$ is an open set containing $y$. Moreover, by its construction, $V$ is also $SI$-open. We also have that for each $i\in\mathbb{N}$, $\twoheaduparrow_{\Irr}z_i\sub ~\uparrow x$. Hence we have that $y\in V\sub\intr_{SI}(\uparrow x)$.

Let $U$ be open in $X$ and $y\in U$. By assumption, we have $\bigsup \thdai y=y\in U$. Since $U$ is inaccessible by suprema of irreducible sets, there exists $x\in X$ such that $x\lli y$ and $x\in U$. By the above result, we have that $y\in \intr_{SI}(\uparrow x)$. Therefore $X$ is a C-space.
\qed

The condition that $X$ satisfies $\oplus$-property in Lemma \ref{lem:kbsober+irr+implies_Cspace} is essential as witnessed by the space given in Example \ref{ex:cofinite_topology}. The following theorem is an immediate consequence of Lemma \ref{lem:SI(X)_is_C_then_X_is_Irr+}, Lemma \ref{lem:kbsober+irr+implies_Cspace}, and \cite[Theorem 6.4]{zhaoho15}.

\begin{thm}
Let $X$ be a sup-sober space. Then the following conditions are equivalent:
\begin{enumerate}
\item $X$ is $\Irr^+$-continuous.
\item $X$ is a C-space.
\end{enumerate}
\end{thm}

\begin{cor}
In the presence of sup-sobriety, $\SI$-continuity and $\Irr^+$-continuity are the same notion. 
\end{cor}

\section{Convergence class defined by irreducible sets}
\label{sec: convergence class defined by irreducible sets}
In a topological space, approximation can be described by means of net convergence.
Let $X$ be a set. 
A \emph{net} $(x_i)_{i \in I}$ in $X$ is a mapping from a directed set $(I,\leq)$ to $X$, where $\leq$ is a pre-order on $I$.
Real number sequences, for instance, are nets in the Euclidean space $\mathbb{R}$.  Thus, nets can be viewed as generalised sequences.  We denote the class of all nets in $X$ by $\Psi X$.

For each $x \in X$, one can define a \emph{constant net} by $x_i = x$ for all $i \in I$.  Parallel to the notion of subsequence, we have the notion of a subnet.
A net $\left(y_j\right)_{j\in J}$ is a \emph{subnet} of $\left(x_i\right)_{i\in I}$ if (i) there exists a function $g:J \to I$ such that $y_{j} = x_{g(j)}$ for all $j \in J$ and (ii) for each $i \in I$ there exists $j^\prime \in K$ such that $g(j)\geq i$ whenever $j\geq j^\prime$.

A convergence class $\mathcal{S}$ in a set $X$ is a relation between $\Psi X$ and $X$.
An element of $\mathcal{S}$ is denoted by $((x_i)_{i \in I},x)$ or sometimes
$(x_i)_{i \in I} \xrightarrow{\mathcal{S}} x$, in which case we say that the net
$(x_i)_{i \in I}$ \emph{$\mathcal{S}$-converges} to $x$.

Every space $(X,\tau)$ induces a convergence class
$\mathcal{S}_\tau$ defined by
\[
(x_i)_{i \in I} \xrightarrow{\mathcal{S}_\tau} x \iff
\forall U \in \tau.~(x \in U) \implies x_i \in U \text{ eventually.}
\]
Here, a property of a net $(x_i)_{i \in I}$ holds eventually if there exist $i_0 \in I$ such that for all $i \geq i_0$, the property holds for $x_i$.

Given a set $X$ and a topology $\tau$ on $X$, when $(x_i)_{i \in I} \xrightarrow{\mathcal{S}_\tau} x$, we say that \emph{$(x_i)_{i \in I}$ converges to $x$ with respect to topology $\tau$}.  A convergence class $\mathcal{S}$ in a set $X$, is said to be \emph{topological} if there is a topology $\tau$ on $X$ that induces it, i.e., $\mathcal{S} = \mathcal{S}_{\tau}$. In fact, for a topological convergence class, the topology inducing it is unique, which is an immediate consequence of the following propostion

\begin{prop}\label{prop:top_containment_char}
Let $X$ be a set and $\tau$ and $\sigma$ be topologies on $X$. Then $\tau\sub \sigma$ if and only if  $\mc{S}_{\sigma}\sub \mc{S}_{\tau}$.
\end{prop}

A special convergence class in a dcpo called the \emph{lim-inf convergence} was first introduced in~\cite{scott72a}. Crucially, this convergence makes use of the directed sets.
It was shown that the lim-inf convergence class in a dcpo is topological if and only if the dcpo is a domain.  Later in~\cite{zhaozhao05}, this lim-inf convergence was modified to create a new convergence class for a general poset.  Recall that in a poset $P$, a net $\kurung{x_i}_{i\in I}$ converges to $y$ provided that there exists a directed subset $D$ of eventually lower bounds of $\kurung{x_i}_{i\in I}$ whose supremum belongs to $\uparrow y$. In that later paper, it was established that the new lim-inf convergence class in a poset is topological if and only if the poset is continuous.

In this paper, we modify the preceding definition of convergence to suit the context of a topological space by replacing the directed subsets with irreducible subsets.
\begin{defi}
\label{def: Irr-convergence}
Let $X$ be a space.  A net $\kurung{x_i}_{i\in I}$ in $X$
is said to \emph{$\Irr$-converge} to $y \in P$ if there exists $E \in \Irr^+(X)$ such that $\bigsup E \geq y$ and for each $e \in E$ there exists $k(e) \in I$ such that for all $i \geq k(e)$ it holds that $x_i \geq e$. An instance of $\kurung{x_i}_{i \in I}$ converging to $x$ is denoted by $\kurung{x_i}_{i \in I} \siconv y$.
\end{defi}
Equivalently, $(x_i)_{i \in I} \siconv y$ if and only if there exists an irreducible subset $E$ of eventually lower bounds of $\kurung{x_i}_{i\in I}$ and whose supremum exists and belongs to $\uparrow y$.

\begin{rem}
In any set, the notions net convergence and filter convergence are equivalent \cite{brunsschmidt55}. In this paper, we prefer to work with the former, different from that in \cite{erne81}. 
Because of this preference, our ensuing development will depends heavily on Kelley's characterisation of topological convergence class \cite{kelley55}.
\end{rem}

For a space $X$, the convergence class in $X$ defined by $\siconv$ is denoted by $\mathcal{I}$.  The rest of this section is completely devoted to studying $\mathcal{I}$ and its relation with sup-sobriety and continuities of spaces.

The following result characterises $\lli$ in terms of the convergence $\siconv$.
\begin{lem}\label{lem: lli char}
Let $X$ be a space, $\kurung{x_i}_{i\in I}$ be a net in $X$, and $y \in X$.
Then $x_i\siconv y$ implies for each $x \in~ \thdai y$, there is $k(x) \in I$ such that for each $i \geq k(x)$ it holds that $x_i \geq x$. Furthermore, if $X$ is either an $\Irr$-continuous or an $\SI^-$-continuous space, then the converse is true.
\end{lem}
\proof
Let $(x_i)_{i \in I} \siconv y$ and $x \lli y$. Then one can find an irreducible set $E$ such that $y \leq \bigvee E$ and for each $e \in E$ there exists $k(e) \in I$ such that $x_i \geq e$ for all $i \geq k(e)$. Using the fact that $x \lli y$, we can find $e_x \in E$ such that $x \leq e_x$. Hence for all $i \geq k(e_x)=:k(x)$ it holds that $x_i \geq x$.

Conversely, if $X$ is $\Irr$-continuous or $\SI^-$-continuous,
there exists an irreducible subset $E$ of $\thdai y$ such that $\bigsup E=y$. The assumption asserts that for each $x\in E$ there is $k(x) \in I $ such that if $i \geq k(x)$ then $x_i \geq x$. Therefore, $(x_i)_{i \in I} \siconv y$.
\qed

From \cite{kelley55}, we know that a convergence class $\mc{S}$ in a set $X$ is topological if and only if it satisfies the following conditions:
\begin{enumerate}[(1)]
\item (Constants). If $\kurung{x_i}_{i\in I}$ is a constant net with $x_i=x$ for all $i$, then $\kurung{\kurung{x_i}_{i\in I},x}\in \mc{S}$.
\item (Subnets). If $\kurung{\kurung{x_i}_{i\in I},x}\in\mc{S}$ and $\kurung{y_j}_{j\in J}$ is a subnet of $\kurung{x_i}_{i\in I}$, then $\kurung{\kurung{y_j}_{j\in J},x}\in \mc{S}$.
\item (Divergence). If $\kurung{\kurung{x_i}_{i\in I},x}\notin \mc{S}$, then there exists a subnet $\kurung{y_j}_{j\in J}$ of $\kurung{x_i}_{i\in I}$ such that for any subnet $\kurung{z_k}_{k\in K}$ of $\kurung{y_j}_{j\in J}$, $\kurung{\kurung{z_k}_{k\in K},x} \notin \mc{S}$.
\item (Iterated limits). If $\kurung{\kurung{x_i}_{i\in I},x}\in \mc{S}$ and $\kurung{\kurung{x_{i,j}}_{j\in J(i)},x_i} \in \mc{S}$ for all $i\in I$, then $\kurung{\kurung{x_{i,f(i)}}_{(i,f)\in I\times M},x} \in \mc{S}$, where $M :=\prod\himp{J(i)\mid i\in I}$.
\end{enumerate}
We shall rely on this result in proving our main result of this paper.

\begin{lem}\label{lem:SI-class_is_topological}
Let $X$ be a space. 
\begin{enumerate}[\em(1)]
\item The convergence class $\mc{I}$ in $X$ satisfies the axioms (Constants) and (Subnets).
\item If $X$ is $\Irr$-continuous or $\SI^-$-continuous, then $\mc{I}$ satisfies the (Divergence) axiom.
\item If $X$ is $\Irr$-continuous and sup-sober, then $\mc{I}$ satisfies the (Iterated limits) axiom.
\item If $X$ is $\SI^-$-continuous, then $\mc{I}$ satisfies the (Iterated limits) axiom.
\end{enumerate}
\end{lem}
\proof\hfill
\begin{enumerate}[(1)]
\item That $\mc{I}$ satisfies the (Constants) axiom is immediate. We now show that
$\mc{I}$ satisfies the (Subnets) axiom.  Let $\kurung{\kurung{x_i}_{i\in I},x}\in \mc{I}$. Then there exists an irreducible subset $E$ of $X$ such that $x \leq \bigsup E$ and for each $e \in E$ there exists $k(e)\in I$ satisfying $x_i \geq e$ for all $i \geq k(e)$.
Let $\kurung{y_j}_{j\in J}$ be a subnet of $\kurung{x_i}_{i\in I}$, with $y_j = x_{g(j)}$ for each $j\in J$. Then there exists $j'(e)\in J$ such that $g(j) \geq k(e)$ whenever $j\geq j^\prime(e)$. Hence for every $j\geq j^\prime(e)$ we have $y_j = x_{g(j)}\geq e$. Therefore, $\kurung{\kurung{y_j}_{j\in J},x} \in \mc{I}$.

\item Suppose $\kurung{\kurung{x_i}_{i\in I},x} \not \in \mc{I}$. By virtue of $X$ being $\Irr$-continuous or $\SI^-$-continuous, 
there exists an irreducible subset $E$ of $\thdai x$ such that $\bigsup E=x$. Hence we can find $y \in E\sub~ \thdai x$ such that for each $i \in I$ one can find $j(i) \in I$ satisfying $j(i) \geq i$ and $x_{j(i)} \ngeq y$.
Define $J := \himp{j \in I \mid x_j \ngeq y}$.
Then $\kurung{x_j}_{j \in J}$ is a subnet of $\kurung{x_i}_{i \in I}$.
For every subnet $\kurung{z_k}_{k\in K}$ of $\kurung{x_j}_{j\in J}$ we have that $z_k \ngeq y$.
By Lemma~\ref{lem: lli char}, $\kurung{\kurung{z_k}_{k\in K},x}$ cannot belong to $\mc{I}$. Thus, $\mc{I}$ satisfies the (Divergence) axiom.

\item We now prove that $\mc{I}$ satisfies the (Iterated limits) axiom.
Let $\kurung{\kurung{x_i}_{i\in I},x} \in \mc{I}$ and
$\kurung{\kurung{x_{i,j}}_{j\in J(i)},x_i} \in\mc{I}$ for all $i \in I$.
Let $y \lli x$. Since $X$ is $\Irr$-continuous and sup-sober, by Theorem~\ref{thm: interpoting prop in SI infty}, the relation $\lli$ is interpolating. Then there exists $z \in X$ such that $y \lli z\lli x$. Applying Lemma~\ref{lem: lli char} to the situation where $x_i \siconv x$ and $z \lli x$, there exists $k(z)\in I$ such that $x_i \geq z$ for all $i \geq k(z)$.  We then have $y \lli x_i$ for all such $i$. Similarly, applying Lemma~\ref{lem: lli char} to the situation where $x_{i,j} \siconv x_i$ and $y \lli x_i$, there exists $g(i) \in J(i)$ such that if $j \geq g(i)$ then $x_{i,j} \geq y$.

Let $M := \prod\himp{J(i)\mid i\in I}$.
Define $h \in M$ such that $h(i) = g(i)$ if $i \geq k(z)$ and $h(i)$ is any element in $J(i)$, otherwise. If $(i,f) \in I \times M$ with $(i,f) \geq (k(z),h)$, then $f(i) \geq h(i) \geq g(i)$, hence $x_{i,f(i)} \geq z$. By Lemma~\ref{lem: lli char}, $x_{i,f} \siconv x$. Therefore, $\mc{I}$ satisfies the (Iterated limits) axiom.

\item The proof is similar to (3) by considering Proposition \ref{prop:interpolation_in_SI*} instead of Theorem~\ref{thm: interpoting prop in SI infty} for the ``the relation $\lli$ is interpolating'' part.
\qed
\end{enumerate}

Lemma \ref{lem:SI-class_is_topological} above provides sufficient conditions for the $\Irr$-convergence class $\mc{I}$ in a space $X$ to be topological.

\begin{lem}\label{lem: subset thadai x}
Let $X$ be a sup-sober space and $x \in X$. If $E$ is an irreducible subset of $X$ such that $\bigsup E \geq x$ and $E \subseteq ~\thdai x$, then $\thdai x$ itself is irreducible in $X$ and has $x$ as its supremum.
\end{lem}
\proof
Let $U_1$ and $U_2$ be open in $X$ such that $\thdai x~ \cap U_1 \neq \emptyset$ and $\thdai x~ \cap U_2 \neq \emptyset$. Then there exist $w_k \in X$ ($k =1,2$) such that $w_k \lli x$ and $w_k \in U_k$.  Since $U_1$ and $U_2$ are upper, we have $x \in U_1 \cap U_2$. Further, since $\bigsup E \geq x$ we have $\bigsup E \in U_1 \cap U_2$. Since $X$ is sup-sober, $U_1 \cap U_2$ is open in $SI(X)$.  This yields that there exists $e \in E$ such that $e \in U_1 \cap U_2$. By assumption, $e \lli x$. Hence $\thdai x \cap U_1 \cap U_2$ is nonempty. We have that $\thdai x$ is irreducible in $X$. Now let $y$ be an upper bound of $\thdai x$. Then $y$ is also an upper bound of $E$. We have that $y\geq \bigvee E\geq x$. Therefore, $\bigvee\thdai x=x$\qed

Lemma \ref{lem: subset thadai x} provides a tool to proof irreducibility of $\thdai x$ in a sup-sober space which was one of our initial intention. 
Bearing in mind that irreducible sets are not necessarily directed (but yet indices of nets are required by definition to be directed sets with respect to some pre-ordering), we are unable to directly deduce $\Irr$-continuity or $\SI^-$-continuity of a sup-sober space from the assumption that the $\Irr$-convergence class in it is topological. However, if we assume further that the space satisfies the $*$-property the $\Irr$-convergence being topological will indeed imply that the space is both $\Irr$-continuous and $\SI^-$-continuous.

\begin{lem}\label{lem:iter_lim_implies_SIcts}
Let $X$ be sup-sober space which satisfies $*$-property. If $\mc{I}$ satisfies the (Iterated limits) axiom then $X$ is both, $\Irr$-continuous and $\SI^-$continuous.
\end{lem}
\proof
Let $x \in X$ and $\mc{F}_x=\himp{\himp{x_{i,j}}_{j\in J(i)}\mid i\in I}$ be the family of all directed subsets of $X$ whose supremum exists and is greater than or equal to $x$. The family $\mc{F}_x$ is nonempty since $\himp{x}$ is in it.

For each $i\in I$, let $x_i := \sup \himp{x_{i,j}\mid j\in J(i)}$.
Then $x_i \geq x$ for all $i\in I$. Since the set $\himp{x} \in \mc{F}_x$, we have $\inf \himp{x_i \mid i\in I} = x$. We define a pre-order $\leq$ on $I$ as follows: $i_1\leq i_2$ for any $i_1,i_2\in I$. We have that $I$ is directed and the net $(x_i)_{i \in I}$ $\Irr$-converges to $x$; just take $\himp{x}$ as the irreducible set satisfying the definition.

For all $i \in I$, define a pre-order $\leq$ on $J(i)$ as follows: $j_1 \leq j_2$ if and only if $x_{i,j_1} \leq x_{i,j_2}$.  We then have $J(i)$ is a directed set and the net $(x_{i,j})_{j\in J(i)}$ $\Irr$-converges to $x_i$; just take $\himp{x_{i,j}\mid j\in J(i)}$ as the required irreducible set.

Let $M := \prod \himp{J(i) \mid i\in I}$. By assumption, we have that the net $\kurung{x_{i,f(i)}}_{(i,f) \in I \times M} \siconv x$.
Thus, we can find an irreducible set $E$ such that
\begin{enumerate}[(1)]
\item $\bigsup E \geq x$ and
\item for each $e \in E$, $x_{i,f(i)} \geq e$ eventually.
\end{enumerate}
We now show that $E \subseteq ~\thdai x$.
Let $e \in E$ and $K$ be an irreducible set with $\bigsup K \geq x$. Since $X$ satisfies $*$-property, there exists a directed set $D$ such that $D\sub ~\downarrow K$ and $\bigsup D=\bigsup K\geq x$. By definition of $\mc{F}_x$, there exists $i_0\in I$ such that $D=\himp{x_{i_0,j}\mid j\in J(i_0)}$. Since $x_{i,f(i)} \geq e$ eventually, there exists $(i_e,f_e)$
such that i $x_{i,f(i)} \geq e$ whenever $(i,f) \geq (i_e,f_e)$.
By the definition of the pre-order defined on $I$, $i_0 \geq i_e$ holds.
Hence $x_{i_0,f_{e}(i_0)} \geq e$.  Since $x_{i_0,f_{e}(i_0)} \in~ \downarrow K$, there exists $k\in K$ such that $x_{i_0,f_{e}(i_0)}\leq k$. It follows that
$e \lli x$. Thus, $E$ is an irreducible subset such that $E \subseteq ~\thdai x$ and $\bigsup E \geq x$, and so by Lemma~\ref{lem: subset thadai x}, $\thdai x$ is irreducible and has $x$ as its supremum. Therefore we have $X$ is $\Irr$-continuous. By Proposition \ref{prop:SI-+kb_implies_*property}, $X$ is also $\SI^-$-continuous.
\qed

Given a convergence class $\mc{S}$ in a set $X$, one defines a topology $\tau_{\mc{S}}$ on $X$ induced by the convergence class, i.e., $U\sub X$ is in $\tau_{\mc{S}}$ if and only if for every $\left((x_i)_{i\in I},x\right)\in \mc{S}$, $x\in U$ implies $x_i \in U$ eventually. From the definition, one can easily see that  $\mc{S}\sub \mc{S}_{\tau_{\mc{S}}}$. The reverse containment $\mc{S}_{\tau_{\mc{S}}}\sub \mc{S}$ is not necessarily true, unless $\mc{S}$ is topological. Indeed, if $\mc{S}$ is a topological convergence class in a set $X$, then the topology on $X$ that induces it is $\tau_{\mc{S}}$ \cite{kelley55}. The following lemma provides the location of the topology $\tau_{\mc{I}}$ on $X$ with respect to the underlying topology and irreducibly-derived topology, assuming that the $\Irr$-convergence class $\mc{I}$ in $X$ is topological.

\begin{lem}\label{lem:location_of_induced_topology}
Let $X$ be a space in which the $\Irr$-convergence class $\mc{I}$ is topological. Then the topology $\tau_{\mc{I}}$ is finer than the irreducibly-derived topology on $X$. If $X$ is $\Irr^+$-continuous or $\SI$-continuous, then the topology $\tau_{\mc{I}}$ is coarser than the underlying topology.
\end{lem}
\proof
Let $x_i\siconv x$ and $U$ be open in $SI(X)$ such that $x\in U$. Then there exists an irreducible set $E$ such that $\bigvee E\geq x$ and for every $e\in E$, $x_i\geq x$ eventually. Upperness of $U$ gives $\bigvee E\in U$. Since $U$ is inaccessible by suprema of irreducible sets, we have $U$ contains an element of $E$ which is an eventually lower bound of the net $(x_i)_{i\in I}$. Hence $\kurung{x_i}_{i\in I}$ converges to $x$ with respect topology on $SI(X)$. Therefore, by Proposition \ref{prop:top_containment_char}, $U$ is in $\tau_{\mc{I}}$.

Now let $(x_i)_{i\in I}$ be a net converging to $x$ with respect to the underlying topology on $X$. If $X$ is $\Irr^+$-continuous or $\SI$-continuous, we are guaranteed to have an irreducible subset $E$ of $\thdai x$ whose supremum is $x$.  For every $g\in E$, $x$ is in the open set $\twoheaduparrow_{\Irr} g$, hence $g\lli x_i$ eventually. This gives that $E$ is a set of eventually lower bound of $(x_i)_{i\in I}$. We have that $(x_i)_{i\in I}$ $\Irr$-converges to $x$. Thus $\tau_{\mc{I}}$ is contained in the underlying topology on $X$, which completes the proof.
\qed

Finally, our main result, i.e., Theorem~\ref{thm: main thm} below, is an immediate consequence of Lemma~\ref{lem:SI-class_is_topological}, Lemma~\ref{lem:iter_lim_implies_SIcts}, Lemma~\ref{lem:location_of_induced_topology}, and Theorem~ \ref{th:SI-infty}.

\begin{thm} \label{thm: main thm}
\begin{enumerate}[\em(i)]
\item
If $X$ is an $\Irr$-continuous sup-sober space, then the net convergence class $\mc{I}$ in $X$ is topological. In addition, if $X$ also satisfies $\oplus$-property, then the topology that induces $\mc{I}$ is exactly the underlying topology on $X$.
\item 
If $X$ is a sup-sober space satisfying $*$-property in which the net convergence class $\mc{I}$ is topological then $X$ is both, $\Irr$-continuous and $\SI^-$-continuous.
\item
A sup-sober space $X$ is $\SI^-$-continuous if and only if it satisfies $*$-property and the net convergence class $\mc{I}$ in it is topological. In addition, if $X$ also satisfies $\oplus$-property, then the topology that induces $\mc{I}$ is exactly the underlying topology on $X$.
\end{enumerate}
\end{thm}

\begin{cor}
In a sup-sober C-space, the topological convergence and $\Irr$-convergence coincide.
\end{cor}


\section{Conclusion}
\label{sec: conclusion}
In this paper, we take a small step towards taking up the programme of exporting domain theory to the more general context of a $T_0$-space.  The key strategy involved in our approach is to simply \emph{replace} directed subsets by irreducible sets -- a methodology first introduced by Zhao and Ho~\cite{zhaoho15}.  Recently, the importance of the role of irreducible (closed) sets in domain theory has also been underscored in the solution of the Ho-Zhao problem in~\cite{hojungxi16}.  All these indicate a need to carry out an in-depth and systematic enactment of the scientific program proposed by Jimmie Lawson (as described in the introduction) via our present \emph{replacement} strategy.  A significant part of our research objective is to see how much of domain theory can be developed in the more general setting of topological spaces.

The main result we report herein characterises those sup-sober spaces satisfying the $\Irr$-continuity (or $\SI^-$-continuity) condition. The fundamental property that sup-sober spaces $X$ are invariant under the Scott derivative operator $SI$ plays a key role in the many major arguments employed herein.  The requirement of sup-sobriety seems indispensable in view that sets of the form $\dua x$ need not be $\tau$-open in an $\Irr$-continuous or $\SI^-$-continuous space $(X,\tau)$.  The present work can be seen as a preliminary investigation of sup-sober spaces which were first introduced in~\cite{zhaoho15}.  We believe that sup-sobriety of spaces is an interesting topic which deserve a more thorough study on its own right.



\begin{thebibliography}{9}
\bibitem{bandelterne83}
H.J. Bandelt and M. Ern\'{e}. The category of $\mc{Z}$-continuous posets. J. Pure Appl. Algebra
30: 219–226, 1983.

\bibitem{baranga96}
A. Baranga. $\mc{Z}$-continuous posets. Discrete Mathematics 152: 33--45, 1996.

\bibitem{brunsschmidt55}
G. Bruns and J. Schmidt. Zur \:{A}quivalenz von Moore-Smith-Folgen und filtern. Mathematische Nachrichten 13: 169--186, 1955.

\bibitem{erne05}
M. Ern\'{e}. Minimal bases, ideal extensions, and basic dualities. Topology Proceedings, 29(2):445--489, 2005.

\bibitem{erne81}
M. Ern\'{e}. Scott convergence and Scott topology in partially ordered sets II. In Continuous Lattices. Vol. 871, pages 61--96. Springer-Verlag Berlin, Heidelberg, New York, 1981.

\bibitem{erne99}
M. Ern\'{e}. $\mc{Z}$-continuous posets and their topological manifestation. Appl. Cat. Structures, 7(1):31--70, 1999.

\bibitem{gierzetal83}
G. Gierz, J.D. Lawson, and A. Stralka. Quasicontinuous posets. Houston Journal of Mathematics, 9:191--208, 1983.

\bibitem{gierzetal03}
G. Gierz, K.H. Hoffman, K. Keimel, J.D. Lawson, M. Mislove, and D. Scott. Continuous Lattices and Domains. Oxford University Press, 2003.

\bibitem{hojungxi16}
W. K. Ho, A. Jung, and X. Xi. The Ho-Zhao Problem. Unpublished note, July 2016.

\bibitem{heckmannkeimel13}
R. Heckmann and K. Keimel. Quasicontinuous Domains and the Smyth Powerdomain. Electronic Notes in Theoretical Computer Science, 298:215--232, November 2013.

\bibitem{johnstone81}
P. T. Johnstone. Scott is not always sober. Lecture Notes in Mathematics, 81:333--334, 1981.

\bibitem{kelley55}
J. L. Kelley. General Topology, volume 27 of Graduate Texts in Mathematics. Springer-Verlag, reprinted edition, 1975.

\bibitem{keimellawson09}
K. Keimel and J. D. Lawson. D-completions and d-topology. Annals of Pure and Applied Logic, 159(3):292--306, June 2009.

\bibitem{mislove99}
M. W. Mislove. Local DCPOs, local CPOs and local completions. Electron. Notes Theor. Comput. Sci., 20, 1999.

\bibitem{scott72a}
D. S. Scott. Continuous lattices. In F.W. Lawvere, editor, Toposes, Algebraic Geometry and Logic, volume 274 of Lecture Notes in Mathematics, pages 97--136. Springer-Verlag, 1972.

\bibitem{venogupalan86}
P. Venugopalan. $\mc{Z}$-continuous posets. Houston J. Math. 12: 275–294, 1986.


\bibitem{wrightwagnerthatcer78}
J.B. Wright, E.G. Wagner and J.W. Thatcher, A uniform approach to inductive posets and inductive closure, Theoretical Computer Science 7:55-77, 1978.

\bibitem{xuliu03}
X. Xu and Y. Liu. The Scott topology and Lawson topology on a $\mc{Z}$-quasi continuous domain. Chinese Annals of Mathematics, 24A:3, pages 365-376, 2003.

\bibitem{zhaofan07}
D. Zhao and T. Fan. Dcpo-completion of posets. Theoretic Computer Science, 411(22--24), 2167--2173, 2007.

\bibitem{zhaoho15}
D. Zhao and W. K. Ho. On topologies defined by irreducible sets. Journal of Logical and Algebraic Methods in Programming, 84(1):185--195, 2015.

\bibitem{zhaozhao05}
B. Zhao and D. Zhao. The lim-inf convergence on partially ordered sets. J. Mathematical Analysis and its applications, 309:701--708, 2005.
\end{thebibliography}
\end{document}